# Measurements of Luminescence from Cleaved Silicon

*Dongguang Li*

Edith Cowan University, Australia
d.li@ecu.edu.au

**ABSTRACT**

This paper outlines the results from experiments performed to gain further information about the structure and properties of cleaved silicon surfaces, using vacuum cleavage luminescence detection methods.

The experiments involved detecting the luminescence produced by cleaving thin silicon plates within a high vacuum, by a process of converting the luminescence to an amplified electrical signal.

## 1. INTRODUCTION

It was discovered that single-crystal silicon, Si, when cleaved in a vacuum, emitted light in at least two wavelength regions at low intensities with duration of hundreds of microseconds [1][2].

This paper outlines the results from experiments performed to gain further information about the structure and properties of cleaved silicon surfaces. The experiments involved detecting the luminescence produced by cleaving thin silicon plates within a high vacuum, by a process of converting the luminescence to an amplified electrical signal. The experiments were based on the assumption that surface cleavage and reconstruction may cause electrons to become excited, and the resulting recombination process which involves the combining of an electron in the conduction band with a hole in the valence band, would result in an emission of energy that would be detectable. This hypothesis was supported by simple calculations that predicted that every broken atomic bond on a silicon surface should radiate one photon, thus generating a strong detectable emission signal. Assuming the energy would be approximate the bond breaking energy of 2.1 eV, the corresponding wavelength was calculated to be within the visible region of the spectrum at 590 nm.

Silicon has an indirect bulk band gap therefore band edge transitions are indirect and mostly non-radiative. The radiative changes in bond-energies during the surface-atom reconstruction process had been observed and was found to be much less than anticipated, falling instead within the infrared region.

## 2. CLEAVED SURFACE STRUCTURE OF SILICON

Silicon is a prominent material used in semiconductor devices therefore it is important to understand its surface structure and properties. Although the bulk structure is well known, the complex rearrangements of the first few layers of atoms of clean surface structures have been more difficult to determine. Clean surfaces have generally been obtained by using techniques such as high temperature thermal flashing, or ion sputtering followed by thermal annealing in an ultra high vacuum (UHV), but an alternative method has been to use direct sample cleaving.

Haneman [3] outlines various methods used to investigate the microstructures of clean cleaved silicon surfaces, including scanning tunneling microscopy (STM) and low-energy electron diffraction (LEED). Data obtained from these techniques appear to be consistent with buckled dimer structures for the (100) 2x1, 2x2 and 4x2 surfaces, and with the dimer-adatom-stacking fault (DAS) model for the annealed (111) 7x7 surface [3]. However other low-index surface structures of silicon are not as well understood. In particular, the structure of silicon surfaces produced by cleavage in ultrahigh vacuum shows a (2x1) cell, which usually converts upon annealing to an intermediate (5x5) structure at about $370^{o}C$, depending upon the annealing time. This structure then converts to the (7x7) form at around $550 – 600^{o}C$.

Measurements of angle-resolved photoemission [4] and polarized optical reflectivity [5] and absorption [6] were consistent with the surface chain structure model proposed by Pandey [7] to explain the large (0.8 eV) dispersion deduced from the photoemission in the Γ-J direction of the two-dimensional Brillouin zone. This model appeared to be consistent with the STM measurements and with the ion scattering data after buckling modifications were introduced, but it wasn't supported by the LEED measurements [8]. Upon close examination, the Pandey model was found to depend upon an assumed calculation of the presence of very strong shear forces existing within the first few layers for which no origin could be found in the cleavage process. The experimental results gave indication of new bond overlaps forming while sub-surface bonds were rupturing.

Therefore, the Three-Bond Scission (TBS) Model, which possesses the surface chain features required to satisfy the optical data, is proposed. The TBS model assumes that the cleavage passes through the three-bond plane, not through the single-bond (111) plane.

## 3. EXPERIMENTAL DESIGN

As shown in Fig. 1, the specimen was supported in a specimen holder at the center of curvature of an aluminium vacuum chamber, with a detector located 5-8 mm above the specimen. After cleaving, the resultant luminescence signal detected was amplified by a low noise amplifier then fed into a digital storage adaptor (DSA) that was triggered by the emission with a set pretrigger period. By incorporating several choices of time base and sensitivities, the signal was captured and directly plotted using an x-y recorder. A cathode ray oscilloscope (CRO) was used for display, and signals were stored and processed on a computer system (CMP). The system was calibrated by testing with a pulse stroboscope (flash) and a light emitting diode (LED) driven by a waveform generator which imitated one shot light signal.

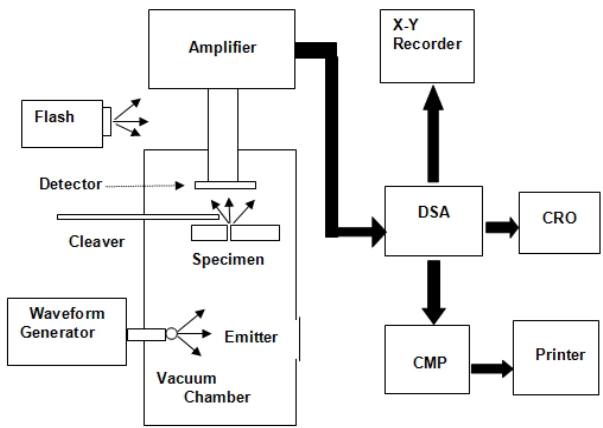

Fig. 1. Block diagram of experimental system

### 3.1. Obtaining High Vacuum

Assuming a unit sticking coefficient, whereby the probability that a molecule striking the surface will stick is 100%, contamination of the sample by a layer of gas forming a compact monolayer (approx. $8 \times 10^{14}$ molec/cm$^2$) occurs at a pressure of $10^{-3}$ Torr in 2.2 ms and at $10^{-6}$ Torr in 2.2 s within an unrestricted supply of ambient gas. These times are extended in systems with volumes of a few litres for surfaces of the order of a few cm$^{-2}$.

Cleavage luminescence usually has the maximum duration of about 1 ms and it is possible to carry out cleavage experiments at high vacuum pressures of $10^{-3}$ to $10^{-6}$ Torr before contamination happens, provided sticking coefficients significantly less than 1 are involved as is usually the case with semiconductor surfaces. A high vacuum condition was achieved by using three different pumps (Rotary, Diffusion and Ion), with the pressure measurable by the current drawn from the ion pump power supply and by an ionization gauge.

### 3.2. Installation of Specimen & Detecting Unit

Cleavage was performed in vacuum which required both specimen and detector to be sealed hermetically in the vacuum chamber, with the specimen fixed on a sample holder combined with a cleaver in the bottom of the vacuum chamber. In room-temperature detectors, an electrical feedthrough mounted on the side of the vacuum chamber could be used for the leading-out of wires of the detector. But the larger liquid-nitrogen-cooled-detectors required a perspex disk of 160cm diameter to replace the chamber cover as a detector holder, with a hole drilled in the center of the holder to fit the outer diameter of the detector case, so that the detector could be located exactly over the sample to be cleaved. This sealing method was used between the detector case and the detector holder as well as between the detector holder and the vacuum chamber. To overcome the mechanical vibration transferred by the metal case which could seriously distort the true signal and reduce the ratio of signal-to-noise, a small size room temperature detector was suspended by fine wires from the Perspex holder over the sample, which effectively obstructed the vibration from the cleaver.

### 3.3. Cleavers

The cleaver design had to reduce the mechanical vibration to a minimum level when the sample was being cleaved. Cleavage can be achieved by bending, pulling or pulling-bending, and several cleaver types were used to allow for several experimental requirements as crack behaviour depended heavily on the cleavage method used.

The block bending cleaver was found to be the best for reducing mechanical vibration. Other types of cleavers used included the paddle bending cleaver, the Wilson seal pulling cleaver, the linear motion feedthrough (LMF) pulling cleaver (Fig. 2) and the paddle pulling-bending cleaver (Fig. 3).

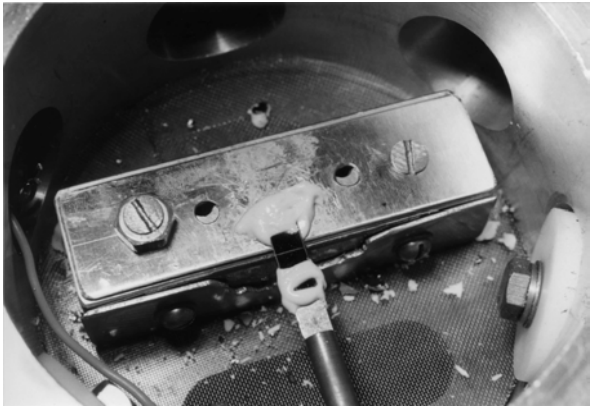

Fig 2. LMF Pulling cleaver

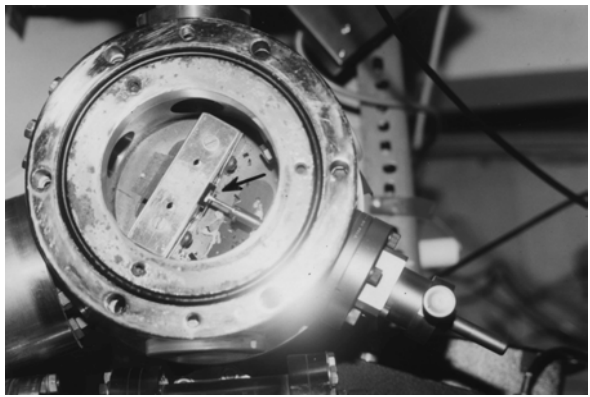

Fig 3. Setting of paddle bending cleaver

### 3.4. Selection of Detectors

Detectors, which convert the luminescence into an amplifiable and measurable electrical signal, are selected according to performance figures of merit based on functions of wavelength and temperature such as responsivity, modulating frequency, bias voltage and the gain of any internal amplifier. The figures of merit considered particularly important were Responsivity (S), Quantum efficiency (n), Response time ($\tau$), Linearity range, Noise equivalent power (NEP), Detectivity (D) and Normalised detectivity ($D^*$).

The detectors were required to detect a weak luminescent signal emanating from any possible reconstruction process and several types of detectors were used because of their varying capabilities. The larger the detector, the easier it was to capture weak signals, but the smaller the detector, the faster the response.

Silicon photodiode detectors were selected to detect larger energy emissions. Their spectral response covers the visible and near infrared 1.1 – 3.1 eV, with excellent linearity and dynamic range. Of the three different Si detectors with differing detecting areas used (1 $cm^2$, 0.36 $cm^2$ and 0.01 $cm^2$), the best was the 1 $cm^2$ photovoltaic unit at room temperature with a detectivity of about $10^{14}$ $cmHz^{1/2}W^{-1}$.

Germanium detectors (photodiodes) were also used to cover the detecting range of 0.67-1.6 eV to increase the spectral sensitivity from 0.8 $\mu m$ to 1.8 $\mu m$, and in particular for their excellent sensitivity for the 800 $nm$ to 1400 $nm$ detection. When cooled to 77°K, they resulted in extremely high shunt impedance for NEP (typically below 0.01 $pW/Hz^{1/2}$) and detectivity above $10^{14}$ $cmHz^{1/2}W^{-1}$. The mercury cadmium telluride (MCT) photoconductive infrared detector was designed for operation in the 1 to 6.5 $\mu m$ wavelength region or 0.19-1.24 eV. The photovoltaic InSb detector with liquid nitrogen cooling generates current when exposed to infrared radiation. One InSb detector used was a high quality indium antimonide photodiode with a detectivity of $10^{11}$ $cmHz^{1/2}W^{-1}$ which provided excellent performance in the 2 to 5.5 $nm$ wavelength range or 0.23-0.62 eV. Single crystal p-n junction technology yielded high speed, low noise detectors with excellent uniformity, linearity and stability. Some of detectors used are shown in Fig. 4.

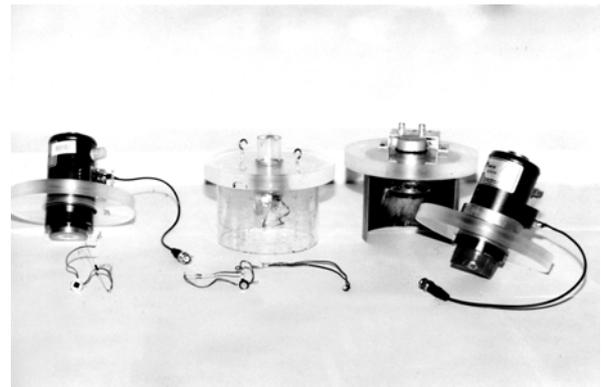

Fig 4. Types of detectors used.
Front L–R: Si (36mm$^2$), Ge (9mm$^2$), Si (1mm$^2$)
Back L-R: Ge (20 mm$^2$); MCT (4mm$^2$) ; Si(100mm$^2$); InSb (13mm$^2$)

## 4. DETACTED SIGNALS

### 4.1. Signal A

Signal A response duration was found to vary from 250 $\mu s$ to 1 $ms$, with many around 800 $\mu s$. There appeared to be at least two components in the emission, shown by the straight line extrapolations in Fig 5 as $T_1$ and $T_2$, with $T_1$ = 358 $\mu s$ (with a S.D. of 84 $\mu s$) and $T_2$ = 783 $\mu s$ (with a S.D. of 270 $\mu s$).

The two or more time constants, $T_1$ and $T_2$, observed for the Si bulk recombination may be related to band structure, different photon cross sections or the nature of the surface as well (tunneling processes). A possible explanation is that after electrons are excited by cleavage from valence band to conduction band, the radiative recombinations occur through several different paths rather than only one process, which results in several overlapped luminescence signals with different time constants.

Experimentally, only a filter with a peak wavelength was 400 nm could effectively block the Si signal, so the conclusion was drawn that the Si signal A has a wide wavelength range of 460-1040 nm (or 1.1-2.7 eV).

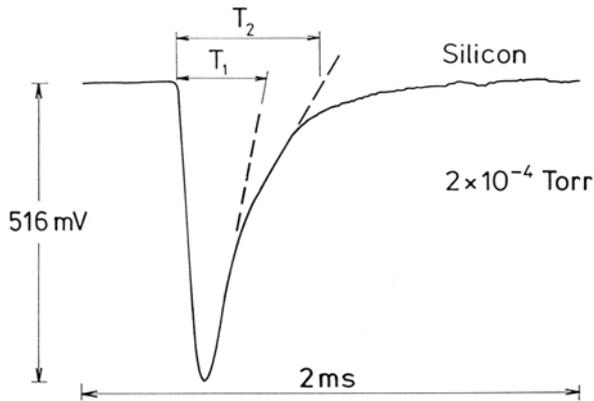

Fig 5. Cleavage luminescence from Si – two time constants

### 4.2. Signal B

The luminescence wavelength from the second cleavage signal, B, corresponds to the recombination of cleavage-excited carriers occurring across the surface state gap. Signal B is harder to detect because the longer wavelength photovoltaic or photoconductive detectors have less detectivity by nearly 3 orders of magnitude than Si photovoltaic detectors. The requirement of keeping the detectors at liquid nitrogen temperature also meant that spurious long-time signals (seconds) were radiated to the detector, caused by vibration of the room temperature parts during cleavage, but these were easily detected and recognized, as indicated in Fig. 6. Problems of background radiation were overcome by using optical interference filters (coarse and segment) mounted directly onto the sapphire window that sealed the vacuum enclosure over the liquid-nitrogen detector. Our experiments then detected the B radiation and showed that the durations were of length of several hundred microseconds, with the average length 200 $\mu s$.

A precise value for the wavelength range of Signal B was obtained from an extensive series of experiments with optical interference filters placed over the liquid-nitrogen-cooled InSb photovoltaic detector.

To determine the energy ranges accurately, the transmittance of Segment 2 with the mask in place was measured with a Cary 17 spectrophotometer. The peak transmittance was 20% with the FWHM (full-width at half maximum) points occurring at 0.266 and 0.253 eV. The points of 10% of peak transmittance occurred at 0.268 and 0.248 eV. Therefore the energy range of Signal B was determined from this to be $0.26 \pm 0.01$ eV.

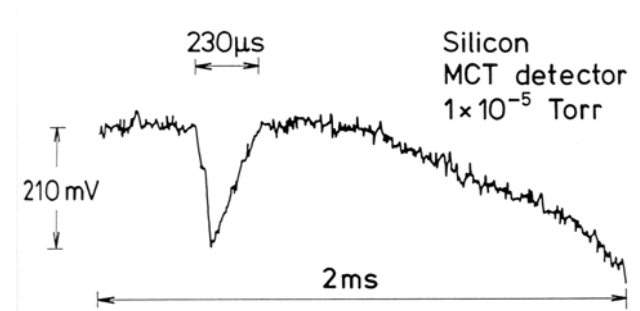

Fig 6. Cleavage luminescence from Si on MCT detector.

### 4.3. Signal C

Signal C was simultaneously found with the usual signal A when faster amplifiers (amplifier B or C) were used, but it is completely different from A in having a duration of approximately 15 $\mu s$ rather than about 300 $\mu s$ for A. Its true duration may in fact be less, as the progress of the crack through the Si wafer would cause continuous emission of light for the duration of the crack.

To detect Signal C, it was necessary to use a very fast amplifier with a sufficiently short response time, although it meant a loss in signal-to-noise performance for any longer signals. A three-resistor-network feedback unit was used to replace the original 10 $M\Omega$ single feedback resistor of the first stage of the amplifier. Signal C was only detectable when cleavage was initiated by longitudinal pulling (i.e. by using the Wilson seal pulling cleaver or LMF pulling cleaver).

Optical filter experiments were carried out to determine the exact wavelength range of the signal C. Among seven optical narrow band-pass filters used, four passed signal C and three blocked it. Experimental results obtained indicated that signal C from cleaved Si also has a wide wavelength spread of 660-1040 nm (or 1.1 – 1.88 eV), compared to signal A which has an even wider wavelength range (0.46 – 1.1 $\mu m$), a characteristic in addition to the difference in duration, which also distinguished signal C from signal A.

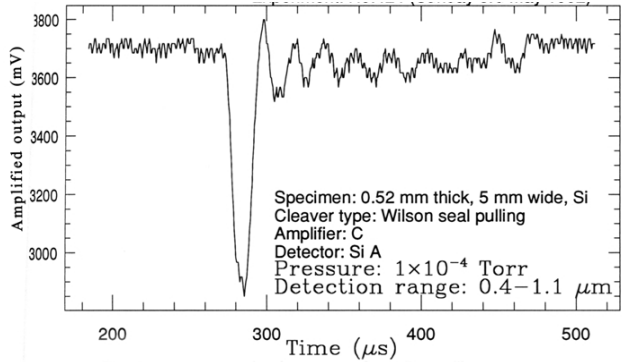

Fig 7. Cleavage luminescence from Si (using pulling cleavers)

## 5. DISCUSSION

The silicon bulk band gap radiation A is explained by assuming that electrons are excited by the cleavage process into the conduction band whence they decay. Since Si has a high density of surface states, there is a strong surface field which would drive conduction electrons into the interior so that their luminescent decay should not be affected by the presence of any non-radiative paths introduced by contamination at the surface. However, the experiments show that the presence of air quashes the radiation, forcing a conclusion that the excited electrons are held at the surface, which could be regarded as a small potential well.

The excited electrons would also occupy surface states. From direct measurements of surface reflectivity [16] and surface absorption [6] on cleaved surfaces with polarized light, it is known that there is a band gap between the surface states with a peak optical-absorption coefficient at approximately 0.45 eV. Hence the observed emission B in the range 0.25-0.27 eV could well be due to transitions across the surface-state gap. Since the duration of B is about 200 $\mu s$, it is inferred that the radiative recombination must be indirect.

In addition to surface state emission, the role of defects has to be considered, since some cleaved Si surfaces display properties that have been ascribed to these (usually interpreted as steps) although other imperfections are also possible. The energy states of such "bad cleaves" have been characterized by optical measurements [16][17], electron energy loss spectroscopy [18][19], pumped photoemission [20][21] and scanning tunneling spectroscopy [22], among others. These agreed in general that there exists a band of defect states with an onset absorption energy at room temperature at about 0.2 eV and a peak at about 0.35 eV. In the case of cleaved surfaces subjected to intense laser irradiation, measurements by Boker et al [20] [21] found short decay times of a fraction of a nanosecond. Very significantly, the lifetimes were shorter for the cases of poorer cleaves.

The position of the defect band with respect to the band edges can be estimated from knowing that the Fermi level is at 0.4 eV above the bulk valence band edge [23], and that the upper surface state band has a peak density at about the Fermi level for the very heavily doped Si (n-type, $8 \times 10^{18}/cm^{-3}$) [14] and about 0.2 eV higher for normally doped material where the band is known to be much less occupied. The lower surface state band has a direct gap from the upper of 0.45 eV. These figures place the top of the lower surface state band at about the top of the bulk valence band. The defect band has a peak at about 0.35 eV above this, i.e. close to the Fermi level.

It is obvious that the energy of 0.26 eV does not fit very well with the peak energies of the defect states (0.35 eV), but could correspond to the onset region of the band, reported as about 0.2 eV [19]. However, if we adopt this interpretation, it must be explained how it is possible to have such a long emission time (200 $\mu s$) from localized defect states. One would have to regard them as trap states with an energy barrier against transition to a state from which fast emission might be possible. From the discussion above of the position of the defect band, the only states that are readily accessible are the upper surface states.

According to the measurements of Bokor et al [20][21] the decay durations of electrons in the upper band are about 2 ns even on defect surfaces. However, in those experiments, the surface was excited into photoemission by a photon pulse of 60 ps duration and power approximately $10^7$ watts $cm^{-2}$. These conditions differ greatly from ours in that the surface is highly excited with a large supply of holes in the valence bonds and of phonons, which are missing in our case. The extra holes and phonons can account for a greatly enhanced transition probability from the upper to the lower surface band. The conditions used by us, in which excitation is due in effect to the energy supplied by lattice vibrations initiated by the bond rupture, will be more characteristic of unexcited Si. We also note that the result (200 $\mu s$) is reasonable for an indirect surface state gap in that long lifetimes are characteristic of device-grade Si, the bulk indirect-gap lifetime being milliseconds.

As discussed previously, the data can thus be interpreted as arising, at least in part, from conduction-band electrons dropping to the surface-state conduction-band minimum and there making transitions to the surface-state valence band. The lowest-energy transitions are indirect, requiring phonons of almost the entire zone-width momentum, and thus accounting for the long observed durations. The data thus supports the feature of the surface state theories in that there is an indirect gap. This feature could not be detected by optical absorption measurements on surface states since it would be too weak.

From the above considerations, the Signal B is explained as due to electrons decaying from the bottom of the surface state conduction band. They reach that band either directly or via defects. In the latter case, the long decay times could be due to trapping times. In the former, they are due to the indirect nature of the surface state gap. In either case, the measured value of 0.26 $\pm$ 0.01 eV would provide a determination of the minimum gap.

With respect to the signal C, there are two related questions: what is the origin of luminescence C and why is it apparently more readily detected during pulling than during bending? At this stage, it is ascribed to recombination at defects on cleaved surfaces since these could have short recombination times. It is known that cleaved surfaces can have, for example, paramagnetic defects in the shape of small partial splits under step edges [24]. Various point defects, mainly vacancies, appear to have been seen in scanning tunneling microscopy study [22][25]. There is as yet, insufficient information to enable one to correlate any particular defect with the mode of cleavage but these results suggest that there is some correlation and that pulling and bending cleavage give noticeably different luminescence results. A microscopic search for topographic differences on many cleavage surfaces formed by the different cleavage techniques failed to find general differences. All surfaces had areas of tear marks, river patterns, and large mirror-like regions. Therefore, we cannot at this stage correlate atomic-scale differences with micron-scale features.